\documentclass[aps,prl,twocolumn,superscriptaddress]{revtex4}
\usepackage{graphicx}
\usepackage{amsmath}
\begin{document}


\title{Direct measurement of the tunneling rate  of the magnetization in 
Fe8 via $^{57}$Fe nuclear spin-lattice 
relaxation by strong collision.}  

\author{S.H.~Baek}
\affiliation{Department of Physics and Astronomy, Iowa State
University and Ames laboratory, Ames, Iowa, 50011}

\author{F. Borsa}
\affiliation{Department of Physics and Astronomy, Iowa State
University and Ames laboratory, Ames, Iowa, 50011}
\affiliation{Dipartimento di Fisica ``A Volta'' e Unita'INFM di Pavia,
via Bassi 6, 271000 Pavia, Italy}
\author{Y. Furukawa}
\affiliation{Department of Physics, Faculty of Science, Hokkaido University, 
Sapporo, 060-0810 Japan}
\author{Y. Hatanaka}
\affiliation{Department of Physics, Faculty of Science, Hokkaido University, 
Sapporo, 060-0810 Japan}
\author{S. Kawakami}
\affiliation{Department of Physics, Faculty of Science, Hokkaido University, 
Sapporo, 060-0810 Japan}
\author{K. Kumagai}
\affiliation{Department of Physics, Faculty of Science, Hokkaido University, 
Sapporo, 060-0810 Japan}

\date{\today}

\begin{abstract}
$^{57}$Fe and $^1$H relaxation measurements have been performed in single 
crystal and oriented powder of enriched $^{57}$Fe8 molecular 
cluster in the temperature range 0.05--1.7 K in zero external field and with 
small perturbing longitudinal field ($< 1$ T).
On the basis of the experimental results it is  
argued  that in zero external field the nuclear spin-lattice relaxation 
($1/T_1$) mechanism is driven by a strong collision mechanism whereby  
$1/T_1$ is a direct measure of the incoherent tunneling probability in the low 
lying magnetic energy states of the molecular nanomagnet.  
The approximate 
value of the effective tunneling rate vs $T$ and $H$ derived directly from 
$1/T_1$ is shown to be consistent with theoretical estimates based on known 
parameters of the Hamiltonian.  

\end{abstract}


\maketitle

Single molecule magnets (SMM) are magnetic systems formed by a cluster of 
transition metal ions within large 
organic molecules \cite{papaefthymiou,gatteschi}. 
SMM are characterized by nearly identical and 
magnetically isolated molecules with negligible 
intermolecular exchange interactions, which allows the investigation of 
nanomagnetism from the macroscopic 
measurement of the bulk sample. Recently,  SMM have been paid much attention 
not only for the fundamental physical properties but also for the potential 
applications in quantum computing and data 
storage \cite{leuenberger2}. Among the molecular magnets, Mn12ac and Fe8 
clusters \cite{lis,delfs}, which have a high total ground state 
spin ($S = 10$), are of particular interest due to the superparamagnetic 
behavior and the quantum tunneling of the 
magnetization (QTM) observed at low temperature 
\cite{thomas,friedman,wernsdorfer1}  due to the large uniaxial anisotropy.
The octanuclear Fe$^{3+}$ cluster \cite{delfs} (Fe8) is a particularly 
good candidate for the study of quantum effects 
since it couples an uniaxial anisotropy leading to an energy barrier of  25 K 
to a non negligible in-plane 
anisotropy. The latter is crucial in enhancing the tunneling splitting of the pairwise degenerate magnetic quantum
states. In fact, Fe8 shows pure quantum regime below 0.4 K \cite{wernsdorfer1}. 
Moreover, it was found that the enrichment of $^{57}$Fe isotope  
in Fe8 shortens the relaxation time demonstrating that the hyperfine field 
plays a key part in QTM \cite{wernsdorfer4}. Together 
with intensive theoretical investigations 
\cite{raedt,boutron,prokofev2,alonso},  QTM in Fe8 has  been 
revealed by various techniques such as
magnetization measurements \cite{wernsdorfer1}, 
ac-susceptibility \cite{sangregorio}, specific heat 
measurement \cite{luis}, 
and nuclear magnetic resonance 
\cite{furukawa03,ueda2,borsa2}.  

Previous proton NMR studies on Fe8 have yielded information about hyperfine 
interaction, fluctuations of the local moments of Fe$^{3+}$ ions 
\cite{furukawa01}, and tunneling effects \cite{ueda2,furukawa03}. More 
recently,   
$^{57}$Fe NMR studies have yielded direct information on the local magnetic 
structure of the ground state and 
the hyperfine interactions \cite{furukawa03r}. 
In the course of the above NMR studies it has become apparent that at low 
temperature the magnetic moment of the Fe8 molecule becomes static in the 
timescale of the NMR experiment 
and thus one can observe both $^1$H and $^{57}$Fe NMR  
in zero external field. In this low temperature region the spin dynamics is 
dominated by incoherent tunneling between pairwise degenerate $m$ magnetic 
states. In a  NMR experiment performed in the local hyperfine field, whenever 
a tunneling event occurs, the quantization axis of the nuclear spins reverses 
its direction and therefore the conventional perturbative approach cannot be 
used to describe nuclear relaxation as first pointed out by Morello 
\cite{morello2}. We thus thought about applying a strong collision theory 
whereby the nuclear $1/T_1$ is predicted to be directly proportional to the 
tunneling rate.  Prompted by this circumstance we have undertaken a 
systematic investigation of the zero field $^{57}$Fe and $^1$H NMR in Fe8 with 
the aim  
of both characterizing this very interesting and seldom observed nuclear 
relaxation regime and of obtaining information about the incoherent tunneling 
rate of the magnetization. 

The formula of the molecular cluster is 
$\mathrm{[Fe_8 (tacn)_6O_2(OH)_{12}]^{8+}[Br_8\cdot9H_2O]^{8-}}$  
(in short Fe8) where tacn is the organic 
ligand 1,4,7-triazacyclonane. Fe8 
consists of eight Fe$^{3+}$ ions ($s = 5/2$) 
whereby the antiferromagnetic interactions among the Fe$^{3+}$ spins lead to a 
total spin $S=10$ ground state \cite{delfs}.

The $^{57}$Fe and $^1$H NMR measurements were performed in both a sample of 
oriented powder and a single crystal both enriched in the $^{57}$Fe isotope. 
The relaxation measurements were 
performed by standard  
Fourier Transform (FT) pulse spectrometers, using a saturating radio 
frequency sequence of several 90$^\circ$ pulses and detection of the nuclear 
magnetization by a 90$^\circ$--180$^\circ$ spin echo sequence. 
The $^{57}$Fe  zero field NMR 
spectrum is composed of eight resonance lines \cite{furukawa03r}. The 
measurements  
presented here refer to the line at about 72.4 MHz for oriented powder and at 
about 65.6 MHz in single crystal.  
The $^1$H zero field NMR spectrum is also structured in a complex way. In the 
proton case measurements at different positions of the spectrum were performed 
namely at 23 MHz and 18 MHz. Temperatures from 1.5 K down to 50 mK were 
obtained using both a closed cycle $^3$He cryostat and a $^3$He--$^4$He 
dilution refrigerator cryostat.

The temperature dependence of the $^{57}$Fe $1/T_1$ in zero external 
magnetic field is plotted in 
Fig.~1. 
The main feature here is 
the $T$-independent plateau reached below 0.4 K, the same temperature at which 
the quantum regime is observed in magnetization measurements 
\cite{wernsdorfer1}. 
In Fig.~2 we show the field dependence at 1.35 K of 
the $^{57}$Fe $1/T_1$ in oriented powder with the magnetic field applied 
along the main anisotropy axis $z$. The main feature in this case 
is the sudden drop of the relaxation rate when a small 
longitudinal field is applied. 
Again, this is a clear indication of the presence of a contribution to 
relaxation due to quantum fluctuations, contribution which is removed by the 
small longitudinal field which prevents quantum tunneling to occur. 

Before analyzing the data in Figs.~1 and 2 quantitatively we discuss the 
mechanism of nuclear spin-lattice relaxation by strong collision. 
In the case of $^{57}$Fe NMR in zero external field and in low fields  a 
tunneling transition between pairwise degenerate states $\pm m$ of the molecule 
results in the rapid change of the local field which is the quantization 
field for the $^{57}$Fe nuclei. In this case a sudden approximation 
strong collision approach should be utilized to describe the nuclear 
spin-lattice relaxation rate. A simple case of strong collision due to a rapid 
inversion 
of a magnetic field is illustrated in Ref.~\cite{abragam}.  The nuclear 
relaxation by strong collision has been treated in details for the case of 
modulation of the nuclear dipolar interaction by ultra-slow diffusional motion 
in insulators \cite{slighter64} and  in the case of quadrupole relaxation by 
a sudden change of the quantization axis as a result of a molecular 
reorientation \cite{alexander}. Since the tunneling event occurs in a time 
much shorter  
than the nuclear Larmor frequency a non adiabatic approach is applicable and 
one has that the nuclear relaxation transition probability $W$ is practically 
the  
same as the tunneling transition probability $\Gamma$ \cite{alexander,abragam} 
i.e.:  
 \begin{equation}
 \label{eq:1}
\frac{1}{T_1} =2W=c(2\Gamma)
 \end{equation}
where $c$ is a constant of the order of one and $\Gamma$ is the effective 
incoherent  
tunneling rate. 
\begin{figure}
\centering
\includegraphics[width=3.2in]{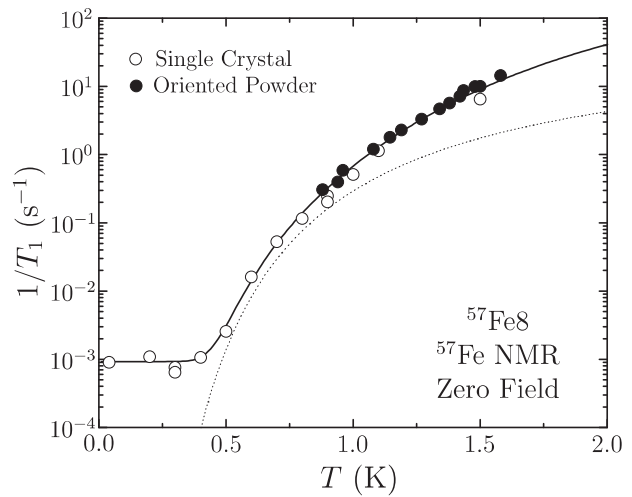}
\caption{\label{fig:fig1}Temperature dependence of $1/T_1$.  
Dotted curve is the calculated spin-phonon contribution based on the 
spin-phonon transition probability \cite{furukawa01,leuenberger}.
Solid curve 
is the sum of the spin-phonon contribution and the tunneling contribution
from Eqs.~(1) and (4) with $c=1$ and the appropriate parameters (see text).}
\end{figure}

On the basis of Eq.~(1) we argue that the measured $1/T_1$ in Fig.~1  
is a direct measurement of the incoherent tunneling probability $\Gamma$.  It 
is instructive to observe that the strong collision result [Eq.~(1)] can be 
obtained as a limit of the weak collision result. In fact, in the weak 
collision case and for random fluctuations of the hyperfine field due to 
incoherent tunneling one has \cite{furukawa03} : 
\begin{equation}
\label{eq:2}
\frac{1}{T_1} = A^2\frac{2\Gamma}{\Gamma^2+\omega_L^2}
\end{equation}
where $A$ is the average fluctuating hyperfine field at the nuclear site and  
$\omega_L$ is the nuclear Larmor frequency. Eq.~(2) was found to describe well 
the proton  
relaxation in Fe8 at high magnetic fields where the tunneling events generate 
a small perturbation of the effective local field i.e.~$A\ll\omega_L$ 
\cite{furukawa03}.  In the limit of slow motion  
($\Gamma\ll\omega_L$) and for the case of a total change of local field 
i.e.~$A\cong \omega_L$ Eq.~(2) 
does indeed reduces to the strong collision case Eq.~(1). 
For the case of $^{57}$Fe NMR the local hyperfine field is directed along the 
magnetization of the molecule and a tunneling event corresponds to a simple  
reversal of the direction of the quantization field. Thus one  may argue that 
$A=\omega_L$ in Eq.~(2) and thus in the slow motion, strong collision limit 
Eq.~(1) and Eq.~(2) coincide for $c=1$. 
On the other hand for $^1$H NMR 
a tunneling event corresponds to a change of both longitudinal and transverse 
components of the local hyperfine field  at the proton site. This may lead to 
$A$ being greater than $\omega_L$ in Eq.~(2) and $c$ greater than one in 
Eq.~(1).  
The  
theoretical estimate of the effect is outside the scope of this paper and thus 
we will treat $c$ as an adjustable parameter in the fit of the proton data 
presented further on.  

In order to support our claim that the nuclear 
relaxation rate is the direct measurement of the tunneling rate we estimate 
the latter on the basis of existing theories and experimental results, and 
compare it to  $1/T_1$. 
\begin{figure}
\centering
\includegraphics[width=3.2in]{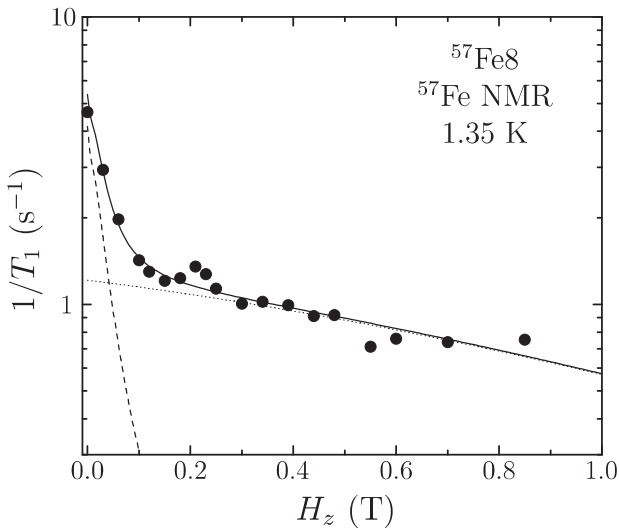}
\caption{\label{fig:fig2}Longitudinal field dependence of $1/T_1$ in oriented 
powder of Fe8.  Dashed line is from Eqs.~(\ref{eq:1}) and (\ref{eq:4}) with 
the same parameters used in the $T$-dependence of $1/T_1$, dotted line is the 
calculated  
spin-phonon  contribution which appears to explain well the high field data, 
and solid line is the sum of the two contributions.  
The small 
enhancement at 0.22 T should be attributed to the level crossing between 
$m=+10$ and $m=-9$ where the tunneling rate increases as the result of the 
pairwise degeneracy of the excited states.}  
\end{figure}
The incoherent 
tunneling probability can be written as \cite{fort,leuenberger,tupitsyn}: 
\begin{equation}
\label{eq:3}
	\Gamma_{m,m'}= \frac{\Delta_{m,m'}^2W_m}{(\xi+\Delta E_{m,-m})^2+W_m^2}.
\end{equation}
$\Delta_{m,m'}$ represents the 
tunneling splitting of the  
corresponding $m$ states. $W_m$ is a broadening parameter of the magnetic $m$ 
state  
which includes both the lifetime broadening due to spin-phonon interaction and 
the hyperfine interaction with nuclei, and $\xi$ is the longitudinal component 
of the internal bias field due to intermolecular dipolar interaction. 
Finally, $\Delta E_{m,m'}$ represents the 
external bias  
due to the application of a longitudinal field which splits the otherwise 
degenerate $m$ states. 
The measured quantity is the 
effective tunneling rate obtained by summing the tunneling probability for the 
different $m$ states weighted by the corresponding Boltzmann factor : 
\begin{equation}
\label{eq:4}
	\Gamma = \sum_m\Gamma_{m,m'}\exp(-E_m/k_BT).
\end{equation}

\begin{figure}
\centering
\includegraphics[width=3.2in]{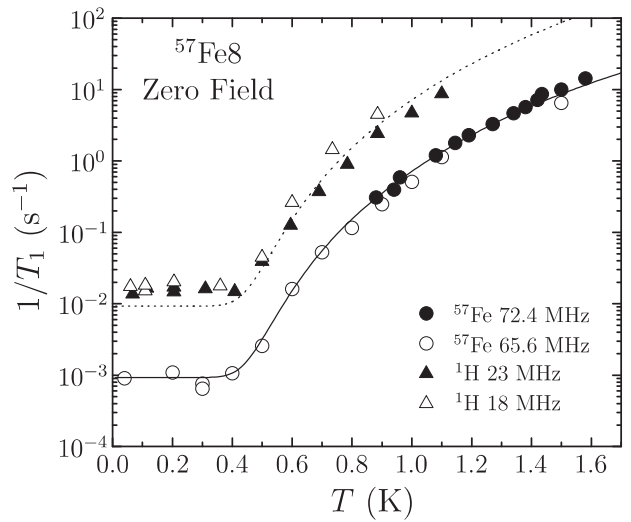}
\caption{\label{fig:3}Comparison of temperature dependence of $^{57}$Fe $1/T_1$
with $^1$H $1/T_1$ at zero field in Fe8. The full line is the same as in 
Fig.~1. The dotted line is the full line multiplied by 10. It is noted that 
the results for both $^1$H and $^{57}$Fe are independent of the Larmor 
frequency $\omega_L$ in agreement with Eq.~(\ref{eq:1}).}
\end{figure}

The 
tunneling splittings necessary to calculate $\Gamma$ from Eqs.~(\ref{eq:3}) 
and (\ref{eq:4}) can be calculated 
from the model Hamiltonian which describes the $S=10$ magnetic ground state 
of the Fe8 molecular cluster:
\begin{equation}
\label{eq:5}
\begin{split}
\mathcal{H} =&\, 
DS_z^2+E\left(S_x^2-S_y^2\right)+\text{g}\mu_B\mathbf{S}\cdot\mathbf{H} 
+D_2S_z^4\\ &+ 
E_2\left[S_z^2\left(S_x^2-S_y^2\right)+\left(S_x^2-S_y^2\right)S_z^2\right]\\ 
&+ C\left(S_+^4+S_-^4\right), 
\end{split}
\end{equation}
where $S_x$, $S_y$ and $S_z$ are the three components of the total spin 
operator, $D$ and $E$ are the axial and the rhombic anisotropy parameter, 
respectively, $\mu_B$ is the Bohr  
magneton, and  the last three terms are the fourth order correction terms. 
The tunneling splitting in the ground state was measured directly with 
Landau-Zener tunneling experiments and found to be $\Delta_{10}\sim 10^{-7}$ K 
\cite{wernsdorfer1}.  
Thus we use in Eq.~(\ref{eq:5}) the values of the parameters $D=-0.293$ K, 
$E=0.047$ K, $D_2=3.54\times 10^{-5}$ K, $E_2=2.03\times 10^{-7}$ K from 
Ref.~\cite{caciuffo} but for $C$ we use a different value i.e., 
$C=-2.7\times 10^{-5}$ K so as to obtain agreement with the experimental value 
of $\Delta_{10}$.  Then by solving Eq.~(\ref{eq:5}) we find 
$\Delta_{10}=0.5\times10^{-7}$ K, $\Delta_9=3.6\times 10^{-6}$ K, and 
$\Delta_8=1.3\times 10^{-4}$ K. With the values of $\Delta_m$ calculated above 
inserted in Eqs.~(\ref{eq:3}) and (\ref{eq:4}) one explains both the field 
dependence in Fig.~2 and the $T$-dependence in Fig.~1 with a choice of fitting 
parameter $W_{10}=2.5\times10^8$ (rad Hz), $W_9=7\times10^9$ (rad Hz), and 
$W_8=9\times10^{10}$ (rad Hz). The parameter $\xi$ in Eq.~(\ref{eq:3}) was set 
$\xi=4.4\times10^9$ (rad Hz) corresponding to the correct order of magnitude 
for intermolecular dipolar fields \cite{prokofev,wernsdorfer3}. The broadening 
parameter  
$W_{10}$ for the ground state is in good agreement with the value measured 
directly by ``hole digging'' experiments \cite{wernsdorfer3}.
The rapid increase of the broadening parameter $W_m$ for $m$ smaller than 10 
is consistent with the rapid increase of the density of states of phonons, 
which contribute to $W_m$ at higher temperatures.
We thus conclude that our measured $\Gamma$ is consistent with the tunneling 
splitting $\Delta_{10}$ and the broadening $W_{10}$ obtained from theory and 
different experiments.  We emphasize once more that only NMR measures directly 
the incoherent tunneling rate $\Gamma$ while the consistency with known values 
of the tunneling splitting is based on Eq.~3 and is thus indirect. 

We compare now the results for $^{57}$Fe NMR with our data for $^1$H  
NMR in Fig.~3. Proton relaxation data in zero field have been published 
earlier \cite{ueda2} in non enriched Fe8. Our data in enriched $^{57}$Fe8 show 
the same $T$-dependence but are almost a factor of two larger which should 
be related to the isotope effect on the tunneling rate 
\cite{wernsdorfer4}.
As can be seen the results for the $T$ 
dependence of the  
proton relaxation in zero field track the ones for $^{57}$Fe with a rescaling 
factor of the order of 10. This is consistent with the argument that the 
relaxation of both nuclei measure directly the effective tunneling rate 
$\Gamma$  
according to Eq.~(1). The multiplication factor of 10 can arise from the value 
of  
the constant $c$ in Eq.~(1) which can be larger for $^1$H NMR for the reasons 
discussed above.

\begin{figure}
\label{fig:fig4}
\centering
\includegraphics[width=3.2in]{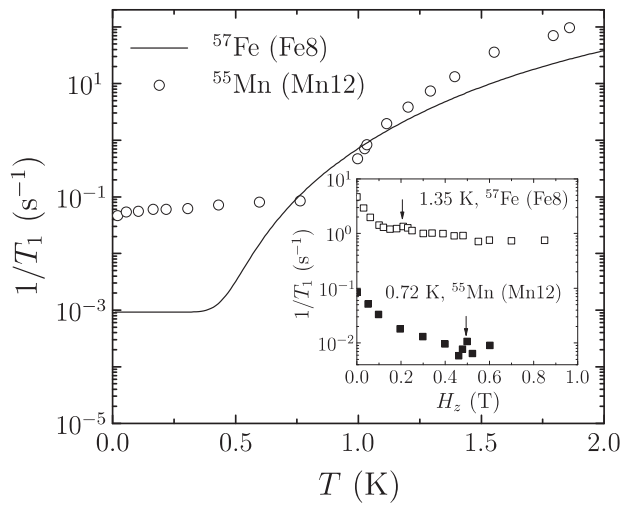}
\caption{Comparison of the temperature dependence of $^{57}$Fe $1/T_1$ in Fe8 
(full line) with 
$^{55}$Mn $1/T_1$ in Mn12, extracted from Refs.~\cite{morello,morello2}. 
Inset shows the analogous comparison for the longitudinal field dependence. 
It is noted that as expected the small anomaly  at the level crossing 
field both in Fe8 ($H_c\sim 0.22$ T) and Mn12 ($H_c\sim 0.5$ T) are observed 
at temperatures where the first excited states are thermally populated, and 
thermal assisted tunneling takes place.}
\end{figure}

Finally the $^{57}$Fe relaxation data in Fe8 are compared in Fig.~4 with the 
$^{55}$Mn relaxation data in Mn12 from Ref.~\cite{morello2}. 
We emphasize that the Mn12 case is more complicate than Fe8. One could 
reinterpret the $^{55}$Mn relaxation data in terms of strong collision with 
a caveat: the $^{55}$Mn $1/T_1$ in Mn12 is dominated by
the presence of a sizeable fraction of fast relaxing molecules combined 
with intercluster nuclear spin diffusion as shown in 
Refs.~\cite{morello,morello2}.  
Thus the low $T$ plateau of $1/T_1$ in Mn12 should not be directly related to 
the  
tunneling rate of the bulk Mn12 sample but rather to a combination of the 
tunneling rate of the fast relaxing molecules and of the intercluster spin 
diffusion rate. 

In conclusion, we have shown that both $^{57}$Fe and $^1$H nuclear relaxation 
in zero external field and at low temperature in molecular nanomagnet Fe8 is 
determined by a strong collision mechanism associated with the reversal of the 
magnetization due to incoherent tunneling. This finding can be generalized to 
other molecules under similar conditions and should thus be very valuable 
because it allows to measure directly through the $1/T_1$ values the effective 
tunneling rate in molecular nanomagnets. 

We thank B.J. Suh for collaboration, A. Cornia for providing the enriched 
powder sample of Fe8, and L.J. de Jongh and A. Rigamonti for useful 
discussions. Ames Laboratory   
is operated for the U.S. Department of Energy by Iowa State University under 
Contract No.~W-7405-Eng-82. This work at Ames Laboratory was supported by the 
Director for Energy Research, Office of Basic Energy Sciences.

\bibliography{fe8PRL}

\end{document}